\documentclass[aps,prl,twocolumn,showpacs,superscriptaddress,groupedaddress]{revtex4}
\usepackage{graphicx}
\usepackage{dcolumn}
\usepackage{bm}

\usepackage{amsmath}

\begin{document}

\title{Unequivocal differentiation of coherent and chaotic light through interferometric photon correlation measurements}

\author{A.\ Lebreton, I.\ Abram, R.\ Braive, I.\ Sagnes, I.\ Robert-Philip, and A.\ Beveratos}

\address{Laboratoire de Photonique et Nanostructures LPN-CNRS UPR-20, Route de Nozay, 91460 Marcoussis, France\\$^*$Corresponding author: isabelle.robert@lpn.cnrs.fr }

\begin{abstract}
We present a novel experimental technique that can differentiate unequivocally between chaotic light and coherent light with amplitude fluctuations, and thus permits to characterize unambiguously the output of a laser.
This technique consists of measuring the second-order intensity cross-correlation at the outputs of an unbalanced Michelson interferometer. 
It is applied to a chaotic light source and to the output of a semiconductor nanolaser whose ``standard'' intensity correlation function above-threshold displays values compatible with a mixture of coherent and chaotic light. 
Our experimental results demonstrate that the output of such lasers is not partially chaotic but is indeed a coherent state with amplitude fluctuations.
\end{abstract}


\maketitle

The electromagnetic field inside a laser cavity operating above the stimulated emission threshold is understood to be in a quantum mechanical coherent state, while below threshold the field is ``incoherent'' or chaotic, consisting of a superposition of waves emanating from many independently- and spontaneously-emitting point sources \cite{Glauber, Loudon}.

The coherence of the laser output is usually evaluated through the second-order intensity autocorrelation function of the laser $g^{(2)}(\tau)$. 
Below threshold, $g^{(2)}(0)$ (i.e.\ at zero time-difference) equals 2, characteristic of a chaotic field, while above threshold it takes the value $g^{(2)}(0) = 1$, characteristic of a quantum coherent state or a classical stable wave.
When the pump power is varied across the threshold in conventional lasers, in which the fraction ($\beta$) of spontaneous emission that is coupled to the ``useful'' mode is very small ($\beta \ll 1$), there is an abrupt transition between these two values, thus providing a handle for the identification of the threshold \cite{Rice1994}.
However, experiments measuring $g^{(2)}(0)$ in high-$\beta$ nanolasers have shown that $g^{(2)}(0)$ undergoes a smooth and gradual transition from the value of 2 to 1 \cite{Ulrich2007, Choi2007, Hofmann2000}. 
For pulsed operation, the drop is so gradual that a value significantly above 1 persists even at pumping powers 4 times above threshold \cite{Elvira2011}.
This feature has been interpreted as indicating that the output of these nanolasers is a ``mixture'' of coherent (stimulated) and chaotic (spontaneous) light, and does not become fully coherent until it approaches the value of $g^{(2)}(0)=1$.

However, according to its definition, $g^{(2)}(0)$ measures only the variance of the intensity fluctuations and gives no information as to whether the field under test corresponds to a statistical ensemble of randomly-phased waves or a coherent quantum state with dynamical amplitude fluctuations.
Thus, in conventional low-$\beta$ lasers the value of $g^{(2)}(0)=1$ simply indicates that their output has no intensity fluctuations and this arises because their gain (i.e.\ the number of excited dipoles) is clamped.
The coherent laser field, nevertheless, undergoes phase fluctuations (corresponding to the Schawlow-Townes phase diffusion \cite{Schawlow}) and these give rise to the finite spectral width of the laser.
On the other hand, a coherent laser field may also undergo amplitude fluctuations, for example, because of noise-excited relaxation oscillations \cite{Druten2000, Takemura2012}. 
In high-$\beta$ lasers, the number of photons at threshold (given by $1/\sqrt{\beta}$) and the number of excited dipoles are small, giving rise to ``discretization noise'' which drives relaxation oscillations in the output intensity \cite{Wiersig2009} and causes the value of $g^{(2)}(0)$ to be significantly higher than 1 at pumping powers high above threshold \cite{Elvira2011, LebretonTBP}. 
Thus the value of $g^{(2)}(0)>1$ is not necessarily indicative of the presence of chaotic spontaneous emission in the laser output, but may arise from dynamical fluctuations of the coherent field amplitude.

In this Letter, we present an interferometric photon-correlation technique that can unambiguously discriminate between coherent and chaotic light.
This technique, which was initially developed to study spectral diffusion in molecules \cite{Brokman2006, Coolen2007}, consists of measuring the intensity cross-correlation between the two output ports of an unbalanced Michelson interferometer in the photon counting regime, while averaging over many fringes and over time. 
By use of this technique, we demonstrate that the output of high-$\beta$ nanolasers is indeed a pure coherent state (albeit with amplitude fluctuations), and not a statistical mixture of coherent and chaotic light.
In view of this demonstration, we present first some highlights of the theory that are necessary for a better understanding of the experiments.

\begin{figure}[h!]
   \begin{center}
   \begin{tabular}{c}
   \includegraphics[width=7.6cm]{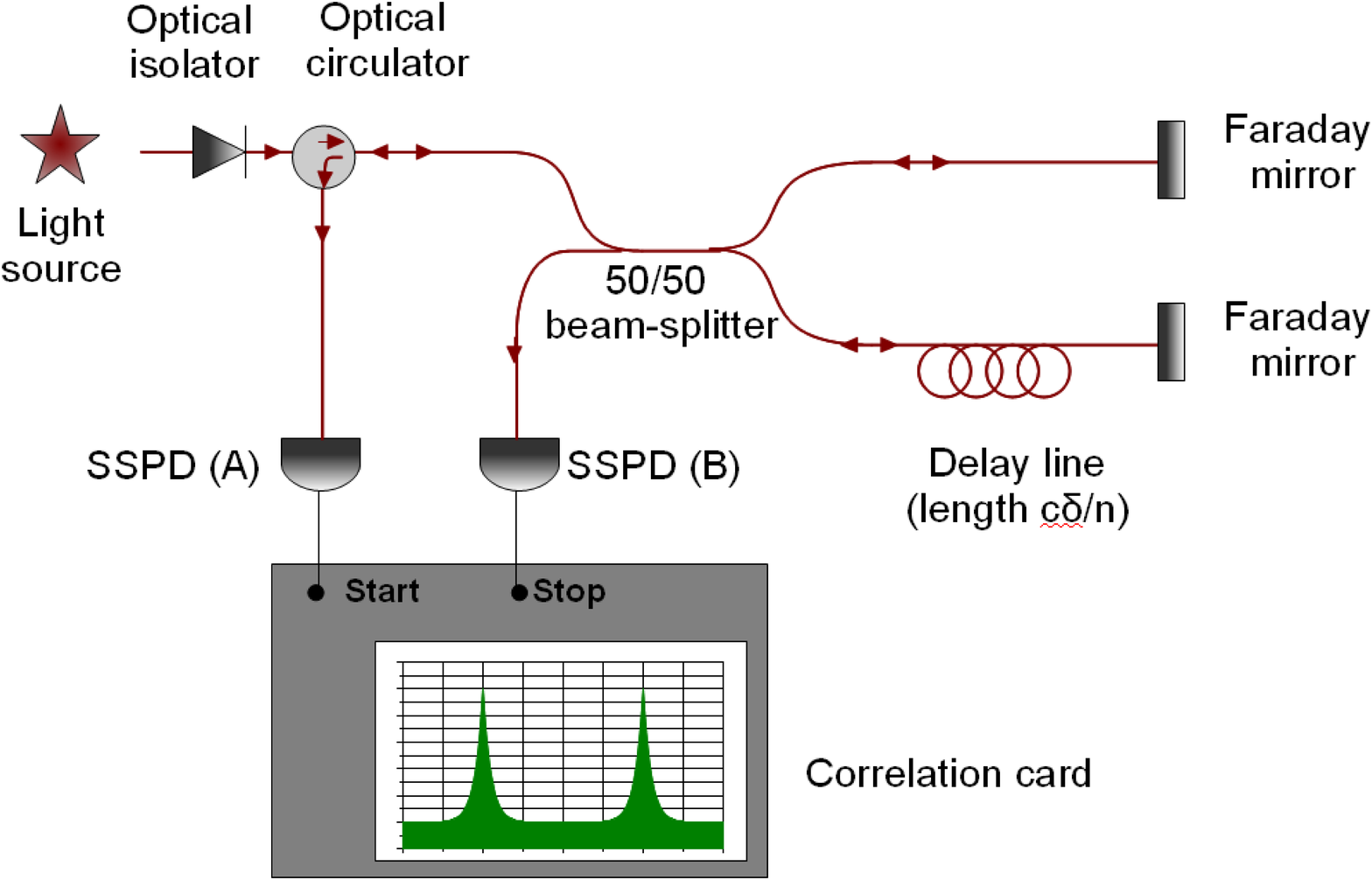}
   \end{tabular}
   \end{center}
   \caption[example]
   { \label{Fig:Setup} Experimental setup. 
   Light is sent into an unbalanced Michelson interferometer made of optical fibers.
   Two single photon detectors (SSPD) measure the cross-correlations between the two output ports of the interferometer.}
\end{figure}

The second-order intensity cross-correlation function between the two outputs of a Michelson interferometer (see Fig.\ \ref{Fig:Setup}) can be written as:
\begin{equation}
g^{(2X)}(\tau,\delta) = \langle E^*_A(t)E^*_B(t+\tau)E_B(t+\tau)E_A(t) \rangle_{t}
\label{eq:g2x}
\end{equation}
where $\tau$ is the time-difference between the two photon detection events, $\delta$ is the ``interferometric delay'' corresponding to the path difference of $\delta c/n$ (where $c$ is the speed of light in vacuum and $n$ is the refractive index of the fiber) between the two interferometer arms, $E^*_i / E_i$ is the positive/negative frequency part of the electric field operator at output port $i=A/B$, $\langle ... \rangle_t$ denotes a quantum mechanical and statistical average, as well as an average over time $t$, while the normalization of $g^{(2X)}$ through a denominator equal to the product of the two output mean intensities is implicit.
The output field may be written in terms of the input field operators,
\begin{eqnarray}
E^*_A(t) & = & \frac{ a^\dagger (t+\delta)- a^\dagger (t)}{\sqrt2}
 \nonumber \\ 
E^*_B(t) & = & \frac{ a^\dagger (t+\delta)+ a^\dagger (t)}{\sqrt2}
\end{eqnarray}
where $a^\dagger (t)$ is the creation operator for the input mode.
An equivalent pair of equations may be written for $E_A/E_B$ and the annihilation operators.
Eq.\ (\ref{eq:g2x}) can thus be expanded into sixteen terms, among which 10 terms average out to zero when the interferometer arm difference $\delta c/n$ is dithered over a distance of a few wavelengths. The six remaining terms are:
\begin{eqnarray}
g^{(2X)}(\tau,\delta) =  \frac{1}{4} & \langle  & a^\dagger (\delta) a^\dagger (\delta+\tau) a(\delta+\tau) a(\delta) 
 \nonumber \\ 
 &+& a^\dagger (0) a^\dagger (\tau) a(\tau) a(0) 
 \nonumber \\ 
 &+& a^\dagger (0) a^\dagger (\delta+\tau) a(\delta+\tau) a(0)
 \nonumber \\ 
 &+& a^\dagger (\delta) a^\dagger (\tau) a(\tau) a(\delta)
 \nonumber \\ 
 &-& a^\dagger (\delta) a^\dagger (\tau) a(\delta + \tau) a(0)
 \nonumber \\ 
&-& a^\dagger (0) a^\dagger (\delta+\tau) a(\tau) a(\delta) \rangle
\label{eq:6terms}
\end{eqnarray}

The first two terms correspond each to $g^{(2)}(\tau)$, the ``standard'' second-order autocorrelation function of the input field, and arise from the two photons propagating like particles along the same interferometer arm, while the third and fourth terms correspond respectively to $g^{(2)}(\tau+\delta)$ and $g^{(2)}(\tau-\delta)$ and arise from the two photons propagating like particles along different interferometer arms. 
These four terms are sensitive only to the intensity fluctuations of the field.
The last two terms, which have a negative sign, describe the fluctuations of the field when it undergoes interference by propagating through both arms of the interferometer, and are sensitive to both amplitude and phase fluctuations.

A simple insight into the shape of $g^{(2X)}(\tau,\delta)$ may be obtained for a strongly unbalanced interferometer in which the delay is much greater than the coherence time of the incoming radiation ($\delta \gg \tau_c$), so that the two peaks corresponding to the displaced autocorrelation functions $g^{(2)}(\tau \pm \delta)$ are far from the region $\tau \approx 0$.
Also, for the sake of simplicity, we assume that the field presents a Lorentzian spectrum of half-width $1/\tau_c$, determined by both amplitude and phase fluctuations. 
We denote the correlation time of the amplitude fluctuations by $\tau_{amp}$. 
At $\tau=0$ the value of the interference term $\langle a^\dagger (\delta) a^\dagger (0) a(\delta) a(0) \rangle$ may be obtained by commuting the two creation operators between them, giving to a good approximation
\begin{equation}
\langle a^\dagger (0)a^\dagger (\delta) a(\delta) a(0) \rangle =g^{(2)}(\delta)\approx 1
\end{equation}
At $\tau>0$ the magnitude of the interference terms decreases.
For a strongly unbalanced interferometer, the correlations dependent on $\delta$ die out, so that for a Lorentzian spectrum we have
\begin{eqnarray}
\langle a^\dagger (\delta) a^\dagger (\tau) a(\delta+\tau) a(0) \rangle & \approx & \langle a^\dagger (\delta) a(\delta+\tau) \rangle \langle a^\dagger (\tau) a(0) \rangle \nonumber \\
& \approx & |g^{(1)}(\tau)|^2= e^{-2|\tau| /\tau_c}
\end{eqnarray}
For a chaotic field, the second-order correlation function obeys the Siegert relation \cite{Loudon}
\begin{equation}
g^{(2)}(\tau)=1+|g^{(1)}(\tau)|^2
\label{eq:Siegert}
\end{equation}
Thus, for a chaotic field the interference terms cancel exactly the Siegert term of $g^{(2)}(\tau)$ and the cross-correlation function in the vicinity of zero time-difference is 
\begin{equation}
g^{(2X)}_{chaotic}(\tau \approx 0,\delta) = 1
\end{equation}
For a coherent field presenting amplitude fluctuations, on the other hand, the second-order autocorrelation function is of the form
\begin{equation}
g^{(2)}(\tau)=1+ \mbox{FT[RIN]}
\end{equation}
where FT[RIN] is the Fourier transform of the Relative Intensity Noise (RIN) spectrum of the laser \cite{Coldren}. 
For a Lorentzian RIN spectrum of width $2/\tau_{amp}$, we have,
\begin{equation}
g^{(2)}(\tau)=1+ \alpha e^{-2|\tau|/ \tau_{amp}}
\label{eq:g2}
\end{equation}
where $\alpha$ is the integral of the RIN spectrum and corresponds to the variance of the fluctuations divided by the square of the mean intensity. 
Thus, for a coherent field the cross-correlation function in the vicinity of zero time-difference is 
\begin{equation}
g^{(2X)}_{coh}(\tau \approx 0,\delta) = 1-\frac12 e^{-2|\tau| /\tau_c} +\frac{\alpha}{2} e^{-2|\tau|/ \tau_{amp}}
\label{eq:g2X}
\end{equation}
Since $\tau_c < \tau_{amp}$, the cross-correlation function displays a dip at $\tau=0$ which goes down to $0.5(1+\alpha)$ and has a width of $\tau_c/2$.
This dip is dug into a broad peak of width $2/\tau_{amp}$, and this is the signature of a coherent field undergoing amplitude fluctuations.
We note that Eq.\ (\ref{eq:g2X}) differs significantly from what is expected for a ``mixture'' consisting of a fraction $x$ of coherent light and $(1-x)$ chaotic light, which is 
\begin{equation}
g^{(2X)}_{mix}(\tau \approx 0,\delta) = 1-\frac{x}{2} e^{-2|\tau| /\tau_c} 
\label{eq:g2Xmix}
\end{equation}

Our experimental setup is described on Fig. \ref{Fig:Setup}. 
The light beam to be tested is introduced in a Michelson interferometer set to a path difference of $\delta c/n$ between its two arms and equiped with photon counting detectors (superconducting single photon counters from SCONTEL) on both of its output ports.
The path difference $\delta c/n$ is dithered over several wavelengths, so that the individual fringes at the output of the interferometer are time-averaged at the detection process. 
The output of the detectors is thus proportional to the fringe contrast.
The outputs of the two detectors are fed in a Becker $\&$ Hickl (B$\&$H) correlation card which plots in real time the normalized photon cross-correlation between the two output ports $g^{(2X)}(\tau,\delta)$ as function of the time-difference ($\tau$) between the two photodetection events, with a resolution of 164 ps.
Polarization alignement of the two arms is achieved by using Faraday mirrors at the ends of the interferometer arms \cite{Zbinden1997}. 
Blocking one arm of the interferometer, allows us to measure directly the standard second-order autocorrelation function $g^{(2)}(\tau)$.

Light from two different sources was studied by means of this technique, so as to have a basis of comparison.
The first source was an unsaturated Erbium Doped Fiber Amplifier with no input signal, thus producing spectrally broad Amplified Spontaneous Emission (ASE) with chaotic statistics \cite{Wong1998}.
The ASE was filtered spectrally by a a fiber Bragg filter (AOS GmbH) centered at 1559.5 nm and having a width of 10 pm, in conjunction with a spectrally broad Yenista tunable filter, thus producing a beam of coherence time $\tau_c\approx 50$ ps \cite{Halder2008}.
The second source was a photonic crystal laser operating at room temperature and emitting at 1601 nm. 
The laser cavity was formed by eleven missing holes in a line (L11) in a photonic crystal consisting of a triangular lattice of holes (diameter 180 nm and period 430 nm) etched in a 260 nm-thick suspended InP membrane. 
The gain material was a single layer of InAsP quantum dots \cite{Michon2008} in the central plane of the membrane. Details on the fabrication of this source can be found elsewhere \cite{Talneau2008}. 
The nanolaser was pumped with a continuous-wave pump at 840 nm, and its threshold was determined  as the inflection point of the logarithmic input-output curve. The laser presented a $\beta$ factor of $\beta \approx 1.5 \times 10^{-3}$, implying that its threshold involves only $1/\sqrt{\beta}\approx 25$ photons.

\begin{figure}[h]
\begin{tabular}{cc}
\includegraphics[width=7.6 cm]{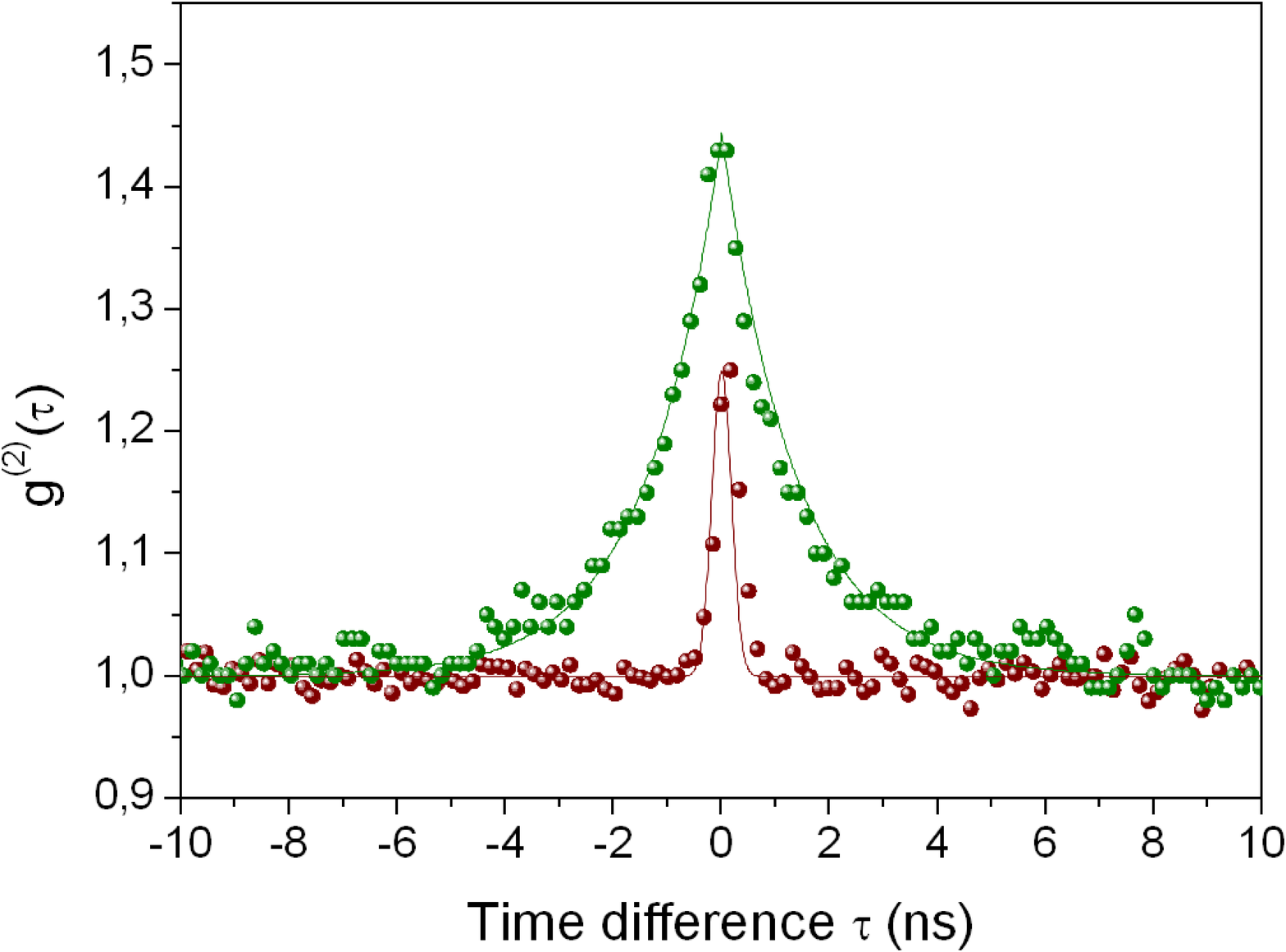} 
\end{tabular}
\caption{Second order autocorrelation function $g^{(2)}(\tau)$ of a nanoscale laser pumped at 1.1 $P_{th}$ (green dots and curve) and a filtered chaotic source (red dots and curve). The points are experimental data and the continuous curves are fits to Eqs.\ (\ref{eq:Siegert}) and (\ref{eq:g2}).}
\label{Fig:g2}
\end{figure}
 
The results of the standard $g^{(2)}(\tau)$ measurements on both sources, are shown on Fig.\ \ref{Fig:g2}. 
The autocorrelation function of the chaotic source (red dots) is independent of pumping power. 
At $\tau=0$ it displays a peak whose height and width are limited by the experimental resolution which is $\sim 4$ times longer than the coherence time (hence $g^{(2)}(0)=1.25$ rather than 2).
The peak can be fitted by a Gaussian (red continuous curve) with standard deviation $\sigma \approx 180$ ps, close to the experimental resolution.
The second-order autocorrelation of the nanolaser, on the other hand, depends on pumping power.
Its value $g^{(2)}(0)=2$ below threshold, drops to $g^{(2)}(0)=1$ at pumping powers of 2 $P_{th}$ (with $P_{th}=$1.3 mW).
The autocorrelation function of the nanolaser, pumped at 1.1 $P_{th}$, is shown on Fig.\ \ref{Fig:g2} (green dots). 
It displays a peak of height $g^{(2)}(0)=1.445$ which drops to $g^{(2)}(\infty)=1$ and can be fitted by an exponential decay (green continuous curve) with time constant $\tau_{amp}/2 = 1.38$ ns, as expected from Eq.\ (\ref{eq:g2}).

\begin{figure}[h]
\begin{tabular}{cc}
\includegraphics[width=7.6 cm]{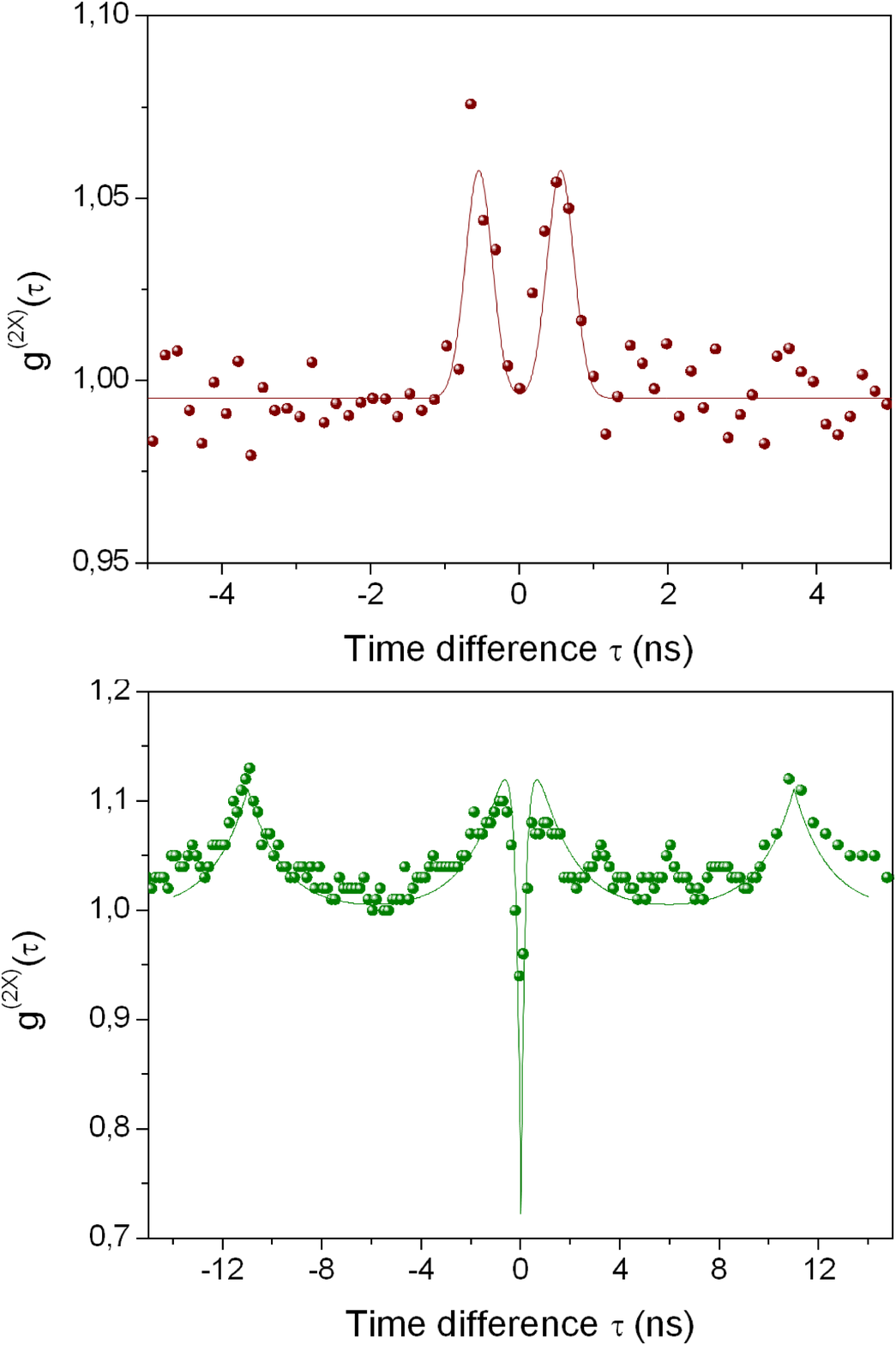}
\end{tabular}
\caption{ Second order cross-correlation function at the outputs of the interferometer. The points are experimental data and the continuous curves are plots of Eq.\ (\ref{eq:6terms}) using the parameters extracted from the fits of $g^{(2)}(\tau)$.
(Top)  Filtered chaotic source. 
(Bottom) Nanoscale laser pumped at 1.1 $P_{th}$.
}
\label{Fig:g2X}
\end{figure}

While both light sources display similar second-order autocorrelation functions which peak at zero time-difference, $g^{(2)}(0)>1$, they exhibit a completely different behavior for their interferometic cross-correlation function, $g^{(2X)}(\tau,\delta)$, as shown on Fig. \ref{Fig:g2X}. 
For the filtered ASE source at $\delta=550$ ps, the cross-correlation function displays two peaks separated in time by $2 \delta$, corresponding to two replicas of the autocorrelation function displaced to $\tau = \pm \delta$ and reduced by a factor of 4, while its value at zero time-difference is $g^{(2X)}(0,\delta)=1$, which is the signature of chaotic light.

For the nanoscale laser, on the other hand, the result is qualitatively different. 
The cross-correlation function at $\delta=11$ ns, in addition to the two replicas of the autocorrelation function at $\tau=\pm \delta$ (as was the case of chaotic light), it displays a strong and narrow dip  at zero time-difference ($\tau \approx 0$) that drops significantly below 1 and is dug into a broad peak with $g^{(2X)}(\tau,\delta)>1$, as expected from Eq.\ (\ref{eq:g2X}). 
This is the signature of a coherent state undergoing amplitude fluctuations.
If the nanolaser output were a ``mixture'' of coherent and chaotic light, the dip would lie on a flat background of $g^{(2X)}(\tau,\delta)=1$ as expected from Eq.\ (\ref{eq:g2Xmix}). 
The experimenal curve is well fitted by Eq.\ (\ref{eq:6terms}) using only the parameters obtained from the fit of $g^{(2)}(\tau)$, namely $\alpha=0.445$, $\tau_{amp}/2=1.38$ ns, and $\tau_c=180$ ps, the resolution of our setup.

The difference between chaotic and coherent light may be understood as follows. 
When a chaotic field is superposed with itself after a delay much longer than its coherence time ($\delta \gg \tau_c$), as the interfering waves in the ensemble have random phases among them, photons may come out from either output port of the interferometer, giving $g^{(2X)}(0,\delta)=1$.
On the other hand, when a coherent field is superposed with itself after a delay of $\delta \gg \tau_c$, even though the value of the phase difference of the two parts will be random (implying that the constructive interference may occur at either one of the two output ports of the interferometer), that value will be maintained for a time of the order of $\tau_c$. Thus, during that time, constructive interference will occur in the same output port of the interferometer (photons will come out from the same port), giving rise to an anti-correlation between the two ports, that is $g^{(2X)}(0,\delta)<1$.
In other words, the unbalanced interferometer, which superposes two time-separated parts of the field, embodies Glauber's view of coherence: ``In physical optics the term is used to denote a tendency of two values of the field at distantly separated points or at greatly separated times to take on correlated values...The coherence conditions restrict randomness of the fields rather than their bandwidth'' \cite{Glauber}.
Thus, measurement of the cross-correlation function at zero time-difference gives an unambiguous indicator of the coherent or chaotic nature of the field.

In conclusion, we proposed and demonstrated an experimental technique based on the second-order cross-correlation of the two outputs of an unbalanced interformeter, which can unequivocally differentiate a chaotic state from a quantum coherent state undergoing dynamical amplitude fluctuations. 
By analysing the output of a high-$\beta$ nanoscale laser through this technique, we have shown that such lasers emit coherent light even when their second-order autocorrelation function at zero time-difference is greater than unity $g^{(2)}(0)>1$, resembling that of a chaotic state. 
This value is not due to the presence of incoherent spontaneous emission in the laser output, but arises from amplitude fluctuations that the coherent laser output undergoes during discrete emission events, in this system of few emitters and few photons.

Acknowledgment: 
The authors acknowledge financial support from the CNANO APOLLON project and from the French National Research Agency (ANR) through the Nanoscience and Nanotechnology Program (project NATIF ANR-09-NANO-P103-36). 
The authors thank H.\ Zbinden, E.\ Diamanti and T.\ Avignon for crucial equipment loans.

\end{document}